\documentstyle{aipproc2}
\begin{document}

\newcommand{\newc}{\newcommand}
\newc{\gsim}{\lower.7ex\hbox{$\;\stackrel{\textstyle>}{\sim}\;$}}
\newc{\lsim}{\lower.7ex\hbox{$\;\stackrel{\textstyle<}{\sim}\;$}}
\newc{\gev}{\,{\rm GeV}}
\newc{\mev}{\,{\rm MeV}}
\newc{\ev}{\,{\rm eV}}
\newc{\kev}{\,{\rm keV}}
\newc{\tev}{\,{\rm TeV}}
\def\vev#1{\left\langle #1 \right\rangle}
\let\p=\partial
\def\beq{\begin{equation}}
\def\eeq{\end{equation}}
\def\bea{\begin{eqnarray}}
\def\eea{\end{eqnarray}}
\def\bi{\begin{itemize}}
\def\ei{\end{itemize}}
\def\benum{\begin{enumerate}}
\def\eenum{\end{enumerate}}
\newc{\ie}{{\it i.e.}}
\newc{\etal}{{\it et al.}}
\newc{\eg}{{\it e.g.}}
\newc{\etc}{{\it etc.}}
\newc{\cf}{{\it c.f.}}
%
%
\let\al=\alpha
\let\be=\beta
\let\ga=\gamma
\let\Ga=\Gamma
\let\de=\delta
\let\De=\Delta
\let\ep=\varepsilon
\let\ze=\zeta
\let\ka=\kappa
\let\la=\lambda
\let\La=\Lambda
\let\del=\nabla
\let\si=\sigma
\let\Si=\Sigma
\let\th=\theta
\let\Up=\Upsilon
\let\om=\omega
\let\Om=\Omega
\def\ph{\varphi}

\title{Classical and Quantum Brane Cosmology\thanks{Plenary talk presented at
CAPP2000, Verbier, Switzerland.}}

\author{John March-Russell}
\address{Theory Division, CERN, CH-1211, Geneva 23, Switzerland}

\maketitle

\begin{abstract}
The first part of this lecture quickly touches upon some important
but infrequently discussed issues in large extra dimension
and warped extra dimension scenarios, with particular reference
to effects in the early universe.
The second part discusses a modification and extension of
an earlier proposal by Brown and Teitelboim to
relax the effective cosmological term by nucleation of fundamental
membranes.
\end{abstract}

\section*{Introduction}

The last two years have seen a ferment in our ideas concerning
beyond-the-standard-model physics and cosmology, the essential new
ingredient being the introduction of the brane concept.
One realization was that in principle, the Standard Model
(SM) could be localized on a brane, or set of branes, embedded in a
higher-dimensional spacetime.  Although such suggestions go back to
the prescient work of Rubakov and Shaposhnikov
\cite{Rubakov:1983bb,Rubakov:1983bz},
the present activity was stimulated primarily by the papers
of Arkani-Hamed \etal
\cite{Arkani-Hamed:1998rs,Antoniadis:1998ig,Arkani-Hamed:1999nn}
(ADD), and Randall and Sundrum \cite{Randall:1999vf,Randall:1999ee} (RS).

ADD showed that the fundamental gravitational
scale (or if string theory is correct the string-scale)
could be close to the TeV scale by invoking large (possibly sub-mm)
new spatial dimensions transverse to the SM brane, implying a very rich
new early universe and collider phenomenology. (In the context of string
theory there were some noteworthy investigations of low-string-scale
or semi-large, $R\sim ({\rm TeV})^{-1}$, extra
dimensions \cite{Antoniadis:1990ew,Lykken:1996fj},
but without the important brane-localization
of the SM degrees of freedom imagined by ADD.)  

An important variant
of the brane scenario involves so-called warped compactifications.  In
such compactifications the directions parallel and transverse to the branes do not
factorize into a product space, leading to non-trivial dependence of the 4D
graviton wavefunction upon the transverse coordinate(s), and possible
localization of the bulk graviton.  Warped (and even non-compact)
``compactifications" were first considered in the 1980's by a number
of authors
\cite{Rubakov:1983bz,vanNieuwenhuizen:1985ri,Nicolai:1985jg,Strominger:1986uh,deWit:1987xg},
but it was only with the addition of the modern brane idea by RS that these speculations
took off.  The phenomenology of these warped models is
in many ways quite different from those of ADD, having typically less
striking signatures at colliders or in astrophysical processes, but concomitantly
being less constrained.
However, in the context of early universe cosmology there have been
quite a large number of papers arguing that the early universe evolution of the RS
models is dramatically different from that of standard FRW cosmology. 
Although, as I will explain, these statements are correct in a certain
high-temperature, high-energy domain, they do not simply apply to sub-TeV
early-universe evolution as many later authors have naively assumed.  In this
talk I will try to disentangle the situation, hopefully for the benefit of those 
freshly arriving (or just interested) in this active field.  

In the second part of my talk I will discuss recent work on some of the quantum
properties of branes and how these might impact upon cosmology.  In this regard
it is important to realize that almost all the brane-world activity (\eg, the ADD
and RS scenarios) take the branes to be {\em classical backgrounds} about which 
a small fluctuation analysis (particle emission for example) is undertaken.  But as
we learned for non-abelian gauge theories there are many significant
features of a theory that are not captured by such a small-fluctuation treatment. 
In particular, I am thinking of instanton processes and false vacuum
``bounce" decays, and monopole and other soliton-like solutions.
Brane-world theories are richer in their dynamics
than non-abelian gauge theories, and one should expect that similarly important phenomena
await investigation and application to the physics of the early universe.  The application
I will describe involves false-vacuum decay enabled by branes
\cite{Feng:2000if}.  As we will see there are properties
of such brane-enabled processes that lead to an essential
alteration of the usual field-theoretic vacuum-decay dynamics, and I will
argue that this could have some bearing on the problem of the cosmological constant
(and possibly lead to novel theories of inflation).
  
\section*{Some aspects of brane cosmology and phenomenology}

The two closely related scenarios of ADD and RS both imagine that
the SM degrees of freedom are localized to
a brane, or set of branes, embedded in an higher-dimensional spacetime.  Here
the branes are taken as a background configuration.
If such a set-up is to be even a crude representation
of our world, then it is vital that we consider a sufficiently rich structure
with the potential to be at least a toy-model for the SM.

As an example of how
this minimal requirement can impact one's view of the physics
of the scenarios, consider the status of the ``solution" of the hierarchy
problem in the warped case. What is usually considered is
a pure gravity set-up with localized brane `sources' of tension, sometimes 
with an additional scalar field in the bulk to provide a stabilization mechanism
if one is concerned with more than one brane.  With the assumption of the bulk
(AdS) cosmological constant and suitably fine-tuned brane tensions one certainly 
finds a metric of the form $d^2s = e^{-2\kappa|y|} \eta_{\mu\nu}dx^\mu dx^\nu
+ d^2y$, with an exponential warp factor.  In the background of this warp factor
it is quite correct to say that all mass scales associated to a theory located at some
$y_c$ in the 5th dimension are conformally rescaled by a factor $e^{-\kappa|y_c|}$.
Does this constitute a solution to the hierarchy problem?  The evidence is no
for the following reason: One must consider the dynamics that localizes
the (toy) SM degrees of freedom on the brane, and in all cases that have been
studied this involves the introduction of further {\em bulk} degrees of freedom.
For example, in string or supergravity investigations of branes in warped
geometries there are always extra degrees of freedom (\eg, the dilaton and other
moduli) who's expectation values at the brane set the values of
coupling constants on the brane.  These additional degrees of freedom
enter the bulk equations in a non-trivial way, and their effect to
turn the exponential warp factor into a polynomial factor
\cite{Chan:2000ms,Mayr:2000zd}, thus eliminating the exponential
hierarchy.

Following Herman Verlinde one argue for this behavior on
general grounds by employing (a generalization of) the AdS/CFT correspondence.
(See section 5 of Ref.~\cite{Chan:2000ms}).  Crudely speaking this
correspondence maps the renormalization group properties of a brane-localized
theory to the geometrical properties of the transverse bulk solution.  The
unperturbed AdS solution corresponds to an exact conformal field theory
on the brane.  But the SM degrees of freedom that we observe are certainly not
conformal; moreover, the SM is unstable to the introduction of relevant
operators (such as the Higgs mass) in the UV.  In the geometrical bulk picture
this instability maps to an instability of the exponential bulk solution to
the couplings of the new bulk degrees of freedom.  The new couplings must be
fine-tuned to get a sufficiently large exponential warp factor, the degree of
tuning being the same as that in the non-susy SM embedded in, say, a GUT
theory.  However, {\em one only sees this instability when the theory is sufficiently
rich to have the potential of realizing the (toy, non-supersymmetric, non-conformal) SM
on the brane.}  Of course if one is happy to have supersymmetry at the TeV scale then
no fine-tuning is needed.  The moral of the story is that some important 
qualitative properties of brane-worlds are very dependent on how simplified a
toy model one is working with.

So forewarned, we now turn to cosmology, and consider first the scenario of ADD.
This case is in part defined by the statement
that the brane-parallel dimensions (for simplicity supposed to be just our usual
3+1 dimensions) and brane-transverse dimensions factorize into a product space. 
\beq
ds^2 = g^{(4)}_{\mu\nu}(x)dx^{\mu}dx^\nu + R^2 g^{(n)}_{ij}(y)
dy^{i}dy^j .
\label{metric}
\eeq
In particular the presence of the brane is assumed not to strongly perturb
the structure of the extra dimensions.  Usually one supposes, following ADD,
that the extra dimensions are close to being flat, and of linear size $R$.
As I will soon explain, this flatness is not a necessary assumption \cite{Kaloper:2000jb}
(for example AdS spaces {\em are} allowed), and relaxing this assumption can,
and does, have a strong effect on the phenomenology. 

The new modes so introduced minimally include at low energy the Kaluza-Klein (KK)
modes of the bulk graviton, while at high energies comparable to the (now much lower
$\gsim \tev$) Planck scale whatever physics renders the theory consistent
-- string excitations of the brane and bulk states being the default assumption.  
The low energy modes are the ones responsible for the stringent astrophysical and 
cosmological constraints that these large extra dimension theories must satisfy,
and of course therefore also have the potential to lead to interesting
predictions.  Before I go on it is worthwhile to
discuss a much more general, and very useful description of the low-energy modes.
Rather like condensed-matter physicists studying a material, let us consider
a plot of the {\em density of states} $\rho(E)$ with
respect to energy of our brane world
scenario.  In the simple case that the extra dimensions are an $n$-dimensional
square torus,
\bea
\rho(E) = & 0 &  \quad E \leq 1/R \nonumber \\
\rho(E) = & {{\rm Area}(S^n)\over 2^n}  R(ER)^{n-1}  &  \quad E > 1/R .
\label{KKspectrum}
\eea
(This ignores the important question
of zero modes to which I will return). 
The Gauss Law constraint, $M_{Pl}^2 = M_*^{(n+2)} R^n$,
relating the traditional Planck scale $M_{Pl}$ to the new fundamental
scale, $M_*$, of gravity implies that there are of order
$M_{Pl}^2/M_*^{2} \sim 10^{32}$ modes in total, so an approximation
by a continuum is good. 

For the large extra dimension scenario we see that the density
of KK states is a power-law function of the mass scale.  The most
important model-independent constraints arise from the production
of light KK modes of the graviton, and so is controlled 
by the low-mass part of this spectrum.  (More precisely for semi-thermal
production from supernovae cores or at the big-bang nucleosynthesis
epoch, various moments of this distribution convolved with a
Boltzmann suppression factor.)
Thus we should ask how much is the shape of the density of states fixed,
and how much is model-dependent?

For geometrical compactifications, these KK modes are understood as the
eigenmodes of the appropriate Laplace operator, $\Delta$, on the internal
space, and all the constraints
depend on the form of the spectral density
of this operator, which in turn depends completely
on the topology and geometry of the internal space.
For example there exist attractive alternate
choices of the compactification implying significantly weaker
constraints~\cite{Kaloper:2000jb}
(admitting in particular a standard 4d FRW evolution
up to quite high temperatures, cf, the usual limit of a few MeV at best
in the $n=2$ flat case, and avoiding the supernovae bounds entirely).
These alternate compactifications
employ a topologically non-trivial internal space,
specifically a $n$-dimensional compact hyperbolic manifold (CHM), with a
large mass-gap to the first non-trivial KK mode.  In other words the low
part of the KK spectrum in Eq.~(\ref{KKspectrum}), is {\em not} model
independent. 
They are also as good, or as bad, as the warped models in
explaining the large hierarchy between the traditional Planck and
weak scales, since even though the volume of these manifolds is large
their linear size $L$ is only slightly larger than the new fundamental
length scale ($L\sim 30 M_*^{-1}$ for example), thusnaively
only requiring numbers of ${\cal O}(10)$ (but beware fine tuning
in sufficiently rich models).

In fact there is no reason why the spectrum of ``KK modes" should
have a geometrical interpretation {\em at all}. (All that is necessary is that a
few sum rules associated to the correct ``running" of the gravitational
interactions are satisfied.  This is so that the gravitational
interaction becomes of ${\cal O}(1)$ strength at the new fundamental
scale of gravity $M_*$.)  This is directly analogous to the asymmetric
orbifold constructions in heterotic string theory which also did not
have a geometrical interpretation of a compactification from 10 to 4
dimensions.  Thus the proper, model-independent way of describing the
physics of, and constraints on, the large extra dimension and warped dimension
models is to introduce the density of states function $\rho(E)$ -- the 
extra-dimensional analog of the parton distribution
functions of the proton!

Finally, what about the possible zero modes I mentioned in passing?  In discussing
the physics of brane-world models one must be careful not to forget such modes
since being massless they can of course have a dramatic effect on the long-wavelength,
low energy behavior of the theory.  A case in point is the radion
mode of the ADD scenario (or the inter-brane separation mode in the original RS scenario).
Before a stabilization mechanism for the size of the extra dimensions is introduced
this mode is massless, and thus the low-energy theory is not pure gravity, but
gravity plus a Brans-Dicke-like scalar.  Such a theory can of course
have very different evolution as compared to the usual Einstein-gravity FRW
solution. Once a stabilization potential is introduced the radion gets a mass, $m$,
and below this mass scale normal FRW evolution is guaranteed by the usual
theorems, anomalous behavior only showing up at energies $E\gsim m$ where
the radion can be excited.  Again the moral of the story is that too
simple a model can lead to misleading conclusions.

I now turn to a discussion of a dynamical process involving branes.

\section*{Brane nucleation and the cosmological term}

An interesting approach to the problem of the cosmological term
is the proposal that it is relaxed by jumps (saltation)
associated with exotic dynamics. There is an important
conceptual advantage to having the relaxation connected to some
non-continuous dynamics that responds only to a source taking
the form of an effective cosmological term.  For if the dynamics
responds to several influences, it is difficult to see how a
particularly simple value for one partial determinant of its behavior
can become overwhelmingly preferred.   Note that this logic also
applies, in connection with {\em continuous\/} relaxation of scalar
fields, to self-interactions, including kinetic terms \cite{Weinberg:1989cp}.

Such saltation of the cosmological term has been previously considered
in the context of stepwise false vacuum decay
\cite{Coleman:1977py,Coleman:1980aw} in a
quasi-periodic staircase potential by Abbott~\cite{Abbott:1985qf}, and
by Brown and Teitelboim (BT) through nucleation of
fundamental membrane degrees of freedom~\cite{Brown:1987dd,Brown:1988kg}.
This second formulation is close in spirit to the one discussed
in \cite{Feng:2000if} and which I review here.

\subsection*{Basic Mechanism of Saltation}

The essential ingredients of the BT model are a 4-form gauge
field strength $F_4 =d A_3$, and a fundamental membrane
degree of freedom (a 2-brane) whose world-volume couples to $A_3$.
Such a field strength $F_4$ has no local dynamics in
4-dimensional spacetime, but its expectation value contributes
to the effective cosmological term.  The presence of an expectation
value for $F_4$ induces the nucleation of the 2-brane to which it
couples, in a manner analogous to the Schwinger mechanism for electric
field decay through nucleation of $e^+ e^-$ pairs.  When a membrane is
nucleated, say as a spherical shell, the effective value of the
cosmological term on the inside differs from the previous value on
the outside, by an amount proportional to the coupling
constant of the membrane.  If a membrane of the correct sign is
nucleated, the contribution of the 4-form to the effective value of
the cosmological term will be reduced. Thus if
the steps between adjacent false vacua are sufficiently small, and if
there is a reason why vacuum decay stops or slows down dramatically as
$\La_{\rm eff} \rightarrow 0$, this mechanism can in principle relax
a large microscopic cosmological term (arising from all sources other
than the membrane-$F$ field dynamics) to a value within observational
bounds. One possible virtue of saltatory relaxation is that, in
contrast to mechanisms involving symmetries or continuous relaxation,
a small non-zero value automatically emerges as a consequence of the
non-vanishing jump size.

Specifically, BT considered gravity in 4 spacetime dimensions
with a 2-brane coupled to a 3-form gauge potential $A_3$. 
The Euclidean space action is
\beq
S_E = \tau_2 \int_\xi \sqrt{\det g_{ab}}
+ \rho_2 \int_\xi A_3 \nonumber \\
- \frac{1}{2} \int d^4 x \sqrt{G} \left(F_4^2 + M^2 (-R + 2 \la)\right)+
{\rm surface~terms},
\eeq
where the $\xi^a$ parameterize the membrane world-volume, and $g_{ab}
= \partial_a X^{\alpha} \partial_b X^{\beta}G_{\alpha\beta}$ is the induced
world-volume metric.  The surface terms 
ensure that the action has well-defined functional derivatives with
respect to the metric and gauge field.  A most important point is that in
four dimensions the 4-form field strength contains no independent
propagating degrees of freedom, its value, up to a constant, being
fully determined by the background of sources charged with
respect to $A_3$.
The parameters entering this action (with 4-d mass dimension $\Delta$)
are: 1) the 2-brane tension $\tau_2, \De=3$; 2) the 2-brane charge density,
$\rho_2, \De= 2$; 3) the Planck mass, $M= (8\pi G)^{-1/2}, \De= 1$; and, 4)
the bare cosmological constant, $\la$, with $\De= 2$. 

The instanton solution is a membrane that divides space into two
regions, an outside $O$ and an inside $I$.  In each region, the field
strength is a constant
$\langle F_4^{O,I}\rangle=c_{O,I}\varepsilon^{(4)}/\sqrt{G}$
and the field strengths are matched across the membrane boundary via
$c_I = c_O - \rho_2$, so that the effective cosmological terms are
\beq
\La_{O,I} = \la + \frac{1}{2 M^2} c_{O,I}^2 \ ,
\label{cceff}
\eeq
where the field strength contribution follows from Einstein's
equations.  {}From Eq.~(\ref{cceff}), it is
clear that if the bare cosmological term is to be canceled, it must be
negative, and therefore $\la<0$ is assumed.

In this BT realization of saltation the tunneling probability
is $P \sim e^{-B}$, where the bounce action for this false vacuum decay is
\beq
B = \left\{ \begin{array}{ll}
\infty\ , & \rm{if}~ \La_O, \alpha_O < 0 \\
12 \pi^2 M^2 \left[ \frac{1}{\La_O} (1-b \alpha_O) -
\frac{1}{\La_I} (1-b \alpha_I) \right] \ , & \rm{otherwise} \ .
\end{array} \right.
\label{bounce}
\eeq
Here the bubble radius, defined so that the area of the bubble slice
when continued back to Minkowski signature is $4\pi b^2$, is
$b = (\La_{O,I}/3 + \alpha_{O,I}^2)^{-1/2}$ with
$\al_{O,I}=\left[c_O-\left(1/2 \pm 3 x^2/4 \right) \rho_2 \right]/3xM$
and $x=\tau_2/\rho_2 M$. (Often Planck
units, $M\equiv 1$, will be used.) The associated change in the effective
cosmological constant across the membrane is given by
\beq
\La_I = \La_O + \frac{1}{2M^2} ( \rho_2^2 - 2 \rho_2 c_O) .
\label{relations}
\eeq
As $\rho_2$ will necessarily be small, while $c_O$ is large so as to (in the end)
cancel the bare cosmological term, the second term inside the
parentheses of Eq.~(\ref{relations}) dominates.

Note that for $\La_O < 0$, transitions can take place only
when $\alpha_O > 0$.  It is not hard to show that this constraint,
along with the reality condition on $b$, implies
that further lowering of the effective cosmological term will not
occur beyond the first anti-de~Sitter step if the tension is
sufficiently large.  This remarkable result is closely related to the
Coleman-de~Luccia~\cite{Coleman:1980aw} gravitational suppression of
false vacuum decay.  BT hypothesize that our Universe is at the
endpoint of such an evolution.

\subsection*{Problems with Abbott-Brown-Teitelboim}

The ABT mechanism is an intriguingly different approach to the
problem of the cosmological term, with a a number of striking
features.  However there are of course a number of
problems to be overcome.  In rough order of increasing seriousness
they are:

\begin{itemize}
\item {\bf Ad-hoc degrees of freedom.}
Dynamical entities are postulated {\em ad hoc},
with no richer context to 
connect them to other principles and problems of physics. 

\item{\bf Small parameters.}
The value of the cosmological term at present isbounded by
$\La_{\rm obs} M^2 \lsim (2\times 10^{-3}\ev)^4 \sim 10^{-120}
M^4$.
For the observed value to be natural in the framework of brane
nucleation, the spacing between allowed values of the effective
cosmological term, near the observed value, cannot be much larger.
This translates into the condition
\beq
\rho_2 \lsim \frac{\La_{\rm obs}}{|\la|^{1/2}} \ .
\label{fineness}
\eeq
This is an extremely stringent condition on the microphysics, even for
plausible $|\la| \ll 1$, since the observed cosmological constant is
so small.  For example, even if the bare cosmological constant is
generated only at the TeV scale through low-energy supersymmetry
breaking, so $|\la| \sim 10^{-60}$, one requires $\rho_2\lsim10^{-90}$. 

Note, however, that if the rest of the ABT mechanism could be made to work
then, even without explaining this very small number, one has
still made considerable progress. The reason for this is that one would
have exchanged a technically unnatural fine-tuning of $\Lambda$, 
for a tiny, but technically natural, coupling $\rho_2$. 

\item{\bf Stopping the saltation.} 
The cosmological term must not only relax to within its observational
bounds, it must also stop evolving once it reaches this interval.
For de~Sitter space, additional bubble nucleations are always
possible.  But in the case of AdS, due to the Coleman-deLuccia
stopping, a sufficient condition to ensure absolute vacuum stability is
$\tau_2^2 >  4\rho_2 c_O/3$ or, in terms of the tension to
charge density ratio,
\beq
x = \frac{\tau_2}{M\rho_2} \gsim \sqrt{\frac{|\la|}{\La_{\rm
obs}}} \ .
\eeq
Unfortunately, even for the smallest plausible $|\la|$, a large hierarchy
between tension and charge density is required if one is to employ
such stopping. 

\item{\bf Penultimate step and an empty universe.}
However more problematic still (and from a phenomenological viewpoint
the most serious defect of the mechanism), is that, as BT showed, upon combining
the stability and step-size conditions, the time required to reach the endpoint is
extraordinarily large, so large that even the glacially slow inflation that
occurs in the penultimate vacuum would leave the universe entirely
devoid of matter and energy.
\end{itemize}


There are a number of possible responses to these problems.  The path advocated
in Ref.~\cite{Feng:2000if} is to consider modifications of the ABT proposal 
motivated by the richer dynamics of the branes of string theory.
Recent developments in string theory have emphasized that a
wide variety of extended objects (which naturally couple to higher-form
gauge fields of the type needed by the BT mechanism) play a fundamental role.
Moreover, such a string theory embedding introduces an essentially new feature,
which may allow a solution of the penultimate step problem of the ABT mechanism.   
The new feature is that there can exist exponentially large
density of states factors associated with semi-classical brane
processes involving coincident branes. This is because such coincident
branes support
low-energy internal degrees of freedom. These degrees of
freedom are not directly visible in a naive semi-classical membrane
calculation (analogous to the microstates
that are deduced to exist for black-holes from the Bekenstein-Hawking
entropy).  The dynamics of relaxation is greatly altered by such large
density of states factors.

In addition since the theory is naturally formulated in higher dimensions, the
properties of these membranes as seen in four dimensions are
determined in terms of the fundamental couplings together with
properties of the extra-dimensional compactification.
Large degeneracy factors and small tension, which give the most rapid
relaxation, appear to arise for complex, near singular configurations. 
Just these configurations are favored for dynamical reasons. Unfortunately
as I will discuss, the simple cases studied so far do not seem to allow
an accommodation of the small coupling necessary for the small step size --
at least not in a theory that contains the SM in a straightforward way.  

Finally, regardless of whether membrane nucleation is the solution,
or perhaps an ingredient, in solving the riddle of the
cosmological constant, nucleation processes involving extended objects
generically occur in string theory, and studying these processes
could lead to insight into the question of vacuum selection.

\section*{(Modified) Saltation Motivated by String Theory}

I shall start with a lightening-quick review of some relevant features of
string/M theory~\cite{JPbook}.
In its long wavelength approximation, M theory supports BPS M2-branes
and M5-branes.  The M2-branes couple
electrically to the 3-index gauge potential $A_3$ of $11$-dimensional
supergravity. Defining a dual gauge potential $A_6$ via
$F=dA_3=*F_7=*dA_6$, one finds that the M5-branes couple magnetically
to $A_3$, or directly to $A_6$.  The
simplest way to arrive at a 4-dimensional world is via
compactification on a 7-dimensional internal space ${\cal{M}}_7$.
(There are also other ways of arriving at a
4-dimensional world involving F-theory or warped solutions that do not
fit into the above framework.  These
may be useful in addressing the step-size problem.)

Two-branes can be obtained from either the fundamental M2-branes,
or by wrapping M5-branes on a 3-cycle $a_3$ of the internal space
${\cal{M}}_7$.
Rather than discuss the properties of such branes in the
general framework of M theory, let's make contact with a possibly
more familiar framework by descending to string theory. 

If ${\cal{M}}_7$ has the form $S^1 \times K$ then for a small $S^1$, M
theory reduces to type IIA string theory, and the M-branes lead to a spectrum
of Dirichlet (D) and Neveu-Schwarz (NS) branes, and gauge fields that couple
to them. Specifically, focus upon the Ramond-Ramond (RR) field strengths
$F_2$ and $F_4$, and in
addition their 8-form and 6-form magnetic duals.  These
$(p+2)$-form RR field strengths couple electrically to dynamical
$p$-branes.

After compactification to (3+1)
dimensions on, say, a Calabi-Yau manifold $K$, the RR fields $F_n$ can
acquire (3+1)-Lorentz invariant expectation values $\vev{F_2} =
v^i_2\, \omega^{(2)}_i$ and $\vev{F_4} = v^i_4\, \omega^{(4)}_i +
v_4\, \ep^{(4)}$, where $\omega^{(2,4)}_i$ are the harmonic 2- and
4-forms on $K$, and $\ep^{(4)}$ is the spacetime volume element.
Denote the expectation values collectively by $v_a$.

Generically, when $v_a\neq 0$, supersymmetry is
broken and a scalar
potential is generated that depends on the Calabi-Yau moduli $T_i$
and the expectation values $v_a$.  The critical points of this scalar
potential are of two general types: either the
Calabi-Yau runs away to infinite volume, or the theory is driven to
conifold-like points where homology cycles of $K$ degenerate
(classically approach zero volume).  Moreover, since the
configurations with $v_a \neq 0$ are not iso-potential, there are
dynamical process whereby the values of the $v_a$'s change.  The $v_a$
`discharge' by the nucleation and expansion of charged membranes, the
D-branes of string theory. What do such false-vacuum decay processes
involving D-branes look like?

\subsection*{Density of states for coincident branes}

As mentioned above, the most important new feature introduced by the 
string theory branes is the existence of density of state factors.
Calculations~\cite{kiritsis} performed in the context of checking
duality between type I and heterotic SO(32) string theory demonstrate
that D-branes do make contributions that can be interpreted
semi-classically as incorporating degeneracy factors reflecting the
non-Abelian structure of coincident D-branes.  Another aspect of this
is that many coincident branes with large total charge can be
described in appropriate limits as `black' objects, similar to black
holes, with event horizons, and with associated Bekenstein-Hawking
(BH) entropy~\cite{KT}.

As a warm-up consider first the case of $k$ coincident D3-branes.  Such a
configuration possesses a U($k$) super Yang-Mills (SYM) gauge theory
on its world-volume.  In the limit where the interactions with the
bulk string theory are weak, and where the temperature (or excitation
energy) of the SYM is small, one can compute the entropy of this
system. When the effective SYM gauge coupling
$g^2_{\rm eff}\simeq kg^2_{\rm YM}$ is small,
the entropy of the gas of massless gauge bosons and
their superpartners at temperature $T$ is simply
\beq
S_3 =\frac{2\pi^2}{3} k^2 V T^3 \ ,
\eeq
where $V$ is the spatial volume of the 3-branes.  What happens when
the effective coupling is large?  In this case one can use the type
IIB supergravity solution describing the $k$ coincident 3-branes.
These classical solutions with the asymptotic geometry and quantum
numbers appropriate for $k$ coincident 3-branes contain a
non-extremality parameter upon which their masses and horizon areas
depend.  If one associates the area of the horizon with BH entropy, one
can derive a temperature by taking an appropriate derivative.  By this
procedure, the supergravity picture yields the strong coupling form of
the entropy.  In this case, the entropy agrees with the preceding weak
coupling formula up to a numerical prefactor 3/4 (which is then a
prediction for the strong coupling behavior of the theory).

For the more interesting case of $k$ coincident D2-branes the
UV theory (in the decoupling limit)
is again a weakly-coupled U($k$) SYM theory, so the UV entropy again
scales as $k^2$.  However one requires the IR entropy since,
as I will soon argue, the physically motivated
temperatures are the ambient de~Sitter temperatures which are small
(vanishingly small as $\La_O \to \La_{\rm obs}$).  In the IR the SYM
theory on the (2+1)-dimensional world-volume becomes strongly coupled,
so one must switch over to the supergravity description. As shown in
Ref.~\cite{malda}, the theory flows in the far IR to that of the M2
brane with BH entropy inferred from the horizon area given
by~\cite{KT}
\beq
S_2 \simeq k^{3/2} A T^2  \ ,
\label{Stwobrane}
\eeq
with $A$ the 2-brane area.  This strongly suggests that such
strongly-coupled brane configurations support ${\cal O}(k^{3/2})$
light degrees of freedom, though the physical nature of these degrees
of freedom remains somewhat mysterious.  
Given this entropy the probability for a
semi-classical process involving $k$ coincident D2-branes is
multiplied by a density of states factor of the form $N_k \sim
\exp\bigl(k^{3/2} A T^2\bigr)$ in the IR limit $T\to 0$.
A more exact treatment requires, among other things,
an additional analysis of the way in which the exponent scales as we
approach the IR, as the physically motivated temperature is small but
non-zero.

In any case, if this reasoning is accepted, the probability for a
semi-classical process involving $k$ coincident 2-branes should be
multiplied by a density of states factor of the form
\beq
N_k \sim \exp\bigl(k^{\be} A T^2\bigr) \ ,
\label{eq:dfactor}
\eeq
for an appropriate temperature $T$.  The exponent $\be$ likely
lies between $\be=2$ and $\be=3/2$.  (Although not utilized 
here, there might also be the possibility of $k^3$ scaling in the
M5-brane limit.)  A larger $\be$ exponent
implies more complete saltation, so $\be=3/2$ is the 
more `conservative' choice.

\subsection*{Temperature ambiguity and a black hole analogy}

The only temperatures intrinsic to our scenario are the de~Sitter
temperatures $T_{O,I}=H_{O,I}/2\pi=(\La_{O,I}/3)^{1/2}/2\pi$.
Ambient ordinary matter might be at a much higher temperature,
but the branes are in very poor thermal contact with it. 
If the initial cosmological term is much larger than the change
brought about by the $k$-bounce,$\La_O \gg k\rho_2 c_O$,
then the de~Sitter temperatures before and after nucleation are almost
identical, $T_O \simeq T_I$, and we may use either one in calculating
the density of states factor.

On the other hand, in the
case that a given transition produces large changes in the
effective cosmological constant, an ambiguity arises.  One possibility
is that the temperature scale for tunneling from a highly curved (high
temperature) de~Sitter space to a less curved de~Sitter space (or even
to a flat or anti-de~Sitter space) is substantially set by the high
temperature.  In this case one would take
$T \sim T_O$ in the density of states factor of Eq.~(\ref{eq:dfactor}).

However, when the change in the nominal de~Sitter temperature is
comparable to the temperature itself, the thermal description of the
tunneling process is internally inconsistent.  A similar situation has
been encountered before, in black hole physics~\cite{sandip}.  The
problem arises in its most acute form for near-extremal holes, as the
temperature approaches zero.  If one uses the temperature of the
initial hole, one finds a significant probability for radiating a
quantum that will take the hole past extremality to a naked
singularity.  A more refined analysis~\cite{KW,kraus} shows that
radiation is {\it not} thermal with regard to the initial temperature,
and in particular that radiation beyond extremality is forbidden.

If an analogy is made between maximally homogeneous cosmologies and
black holes base on their temperatures, then de~Sitter spaces
correspond to ordinary holes, flat space corresponds to an extremal
hole, and anti-de~Sitter spaces to naked singularities.  This analogy
suggests, in view of the previous paragraph, that we should not
consider finite temperature branes that mediate transitions from
de~Sitter to anti-de~Sitter spaces.  As a very crude working hypothesis,
which interpolates smoothly to this suggestion, one can employ the
geometric mean of the de~Sitter temperatures
$T \sim \sqrt{T_O T_I}$, instead of $T\sim T_O$.  

The dynamics of both these possibilities is explored below.

\section*{Two Saltation Scenarios}

Let us now gather the pieces and attempt to envisage if they may
be assembled into a complete scenario.
The cosmological constant evolves from some initial value through
multi-bounce transitions.  The probability for such transitions is
$P \sim e^D e^{-B}$, with the bounce action $B$ is given by
Eq.~(\ref{bounce}), with $\rho_2 \to k\rho_2$ and
$\tau_2 \to k\tau_2$, where $k$ is the bounce number.  The density of
states prefactor is specified by $D=k^{\be} A T^2$, where $A=4 \pi
b^2$ is the 2-brane area, and $T$ is the temperature.  For
concreteness, the low `conservative' value of $\be=3/2$ is assumed,
but simple extensions of the following analysis holds more
generally, for example, for $\be=2$ or larger.  

Since the temperature $T$ is not under good theoretical
control it is useful to explore both of the broad alternatives
mentioned previously.

\subsection*{Single-step relaxation}

First consider the possibility that the temperature is given by the
scale of the initial (outside) de~Sitter temperature, so $T \sim T_O
\sim \sqrt{\La_O}$.
One begins with some bare cosmological constant $\la<0$.  Assume
that the initial field strength gives a similar contribution to the
effective cosmological constant, so $\La_O, c_O^2 \sim |\la|$.
Also assume that some mechanism provides a very small charge
density $\rho_2\sim\La_{\rm obs}/\sqrt{|\la|}$, consistent with the naturality
condition discussed earlier, and that for simplicity $x\simeq 1$.

With such initial conditions and brane properties, the maximal bounce
action is $B \sim 1/|\la|$, while the degeneracy factor may be as
large as $D \sim \la^2 / \La_{\rm obs}^2$.  Recall that
$\La_{\rm obs} \sim 10^{-120}$, while $\la$ is plausibly in
the range of 1 to $10^{-60}$.  Thus, the
degeneracy enhancement overpowers the bounce action suppression, and
tunneling proceeds rapidly.

It is not difficult to show that $D$ is maximized for $k \rho_2 \sim
c_O$, {\em i.e.}, for field strength step sizes of the right order to
neutralize the field strength contribution to the effective
cosmological constant.  Indeed the most probable tunneling events
nucleate bubbles of deep anti-de~Sitter space.  Such events produce
small, short-lived interior universes, so the meaning of `probable' in
this context must be carefully qualified.  Among universes that live a
long time and even remotely resemble ours, the exponentially most
favored are those closest to having zero effective cosmological term.

This scenario invokes a form of the anthropic principle.  It is a
uniquely weak one, however, in the following sense.  Anthropic bounds
on the cosmological term are highly asymmetrical~\cite{Weinberg:1989cp}.
For positive cosmological terms, the formation of sufficiently large
gravitational condensations requires cosmological terms below $\sim
100$ in units of $\rho_c$, the critical density.  For negative
cosmological term, the lifetime of the universe requires cosmological
terms roughly above $-1$.  Thus if the spacing between allowed
near-zero saltatory values of the cosmological term in units of
$\rho_c$ is, say, 3 and allows the values $\ldots, -2, 1, 4, \ldots$,
then among values that can be experienced by sentient observers, 1 is
by far the most likely.

Now suppose such a transition to nearly flat space has occurred.  As
discussed before, absolute stability is possible
only for anti-de~Sitter spaces, and only for extremely large $x$.
However, absolute stability is not required on empirical grounds. One
need only require that the effective value of the cosmological term is
at present stable on cosmological time scales.  For a starting
effective cosmological constant of $\La_O \sim \La_{\rm obs}$, the
degeneracy factor is highly suppressed by small $T$, and the bounce
action is dominant.  The stability of the vacuum is then determined
solely by $B$.

The bounce action can become very large even for almost flat
de~Sitter spaces.  The least
suppressed transition (and so most dangerous from the point of view of
vacuum instability) is that mediated by $k=1$.  For small $\rho_2$ and
small $\La_O$, the $k=1$ bounce action is
\beq
B \approx \frac{27\pi^2 x^4}{2} \left[\frac{\rho_2}{c_O^3}
- \frac{\rho_2^2}{c_O^2 \La_O } \right] \ , \quad
\rho_2, \sqrt{\La_O} \ll c_O \ .
\label{zaza}
\eeq
Neglecting numerical factors, this gives $B\simeq x^4\La_{\rm obs}/\la^2$.
If $|\la| \sim 1$, then the action is very small, and
there is no effective stability.
On the other hand, if supersymmetry is broken at a low scale, then we
expect $|\la|\ll 1$. When does $B\gsim 1$? This translates into
\beq
|\la_{\rm halting}| M^2\lsim x^2 M^2\sqrt{\La_{\rm obs}} 
\simeq x^2 (2\times 10^{-3}\ev)^2 (2.4\times 10^{18}\gev)^2
\simeq x^2\, (2 \tev)^4 \ ,
\label{halting}
\eeq

In a more careful analysis, one may require $B \gg 1$ for stability.
However, the required supersymmetry breaking scale will not differ
significantly from the above estimates, as $B$ goes as the inverse 8th
power of the energy scale appearing in $|\la| M^2$.  In any case, 
given the present experimental lower bounds on the supersymmetry
breaking scale, this suggests that the stability of the vacuum in this
scenario requires low scale supersymmetry breaking, and relates the
cosmological constant, Planck, and weak scales according to
$M_{\rm weak}^2 \sim (10^{-3} \ev) (M_{\rm Planck})$
in accord with observation.

\subsection*{Multi-step relaxation}

Motivated by the black hole analogy, consider now an effective
temperature that is the geometric mean of the initial and final
de~Sitter temperatures.  In this case, tunneling to (and through)
anti-de~Sitter space is forbidden by fiat.  However, the requirements
of rapid tunneling to the observed cosmological constant and its
stability are non-trivial constraints, and we now investigate their
implications.

As in the previous scenario, we consider initial conditions $c_O^2,
\La_O \sim |\la|$.  Now, however, the density of states factor $D$ is
typically maximized for $\La_I$ within an order of magnitude of
$\La_O$.  To see this, a very rough estimate may be obtained by
neglecting the bubble radius dependence on $k$ and approximating
$\La_O - \La_I = (2 \rho_2 c_O - \rho_2^2)/2 \sim \rho_2 c_O$.  We
then have $D \propto \sqrt{\La_O \La_I} (\La_O-\La_I)^{3/2}$, which is
maximized for $\La_I=\La_O/4$.

For $\La_I \sim \La_O$,
\beq
D_{\rm max} \sim \frac{\La_O^{5/2}}{|\la|^{7/4}\rho_2^{3/2}}
\ .
\eeq
It is not hard to verify that this degeneracy factor dominates the
bounce suppression when $\La_O \sim |\la|$.  Thus, initially the
effective cosmological constant tunnels rapidly as in the previous
scenario, but in contrast to the previous case, the cosmological
constant relaxes through several steps, with values roughly following
a geometric series.

The effective cosmological constant will relax as described until
$\La_O \ll c_O^2$, when Eq.~(\ref{zaza}) holds.  At this point,
the condition that tunneling continue is the requirement $D_{\rm max}
\gsim B$, or, since $B \sim \La_O / \la^2$,
\beq
\La_O^{3/2} \gsim |\la|^{-1/4} \rho_2^{3/2} \ .
\label{dmax}
\eeq
For vanishing $\rho_2$, tunneling may continue to arbitrarily small
$\La_O$.  However, if we require stability from $B \gsim 1$, we
find, from Eq.~(\ref{zaza}),
\beq
\rho_2 \gsim c_O^3 \sim |\la|^{3/2} \ ,
\label{rho0}
\eeq
so $\rho_2$ cannot be arbitrarily small.  Combining Eqs.~(\ref{dmax})
and (\ref{rho0}), we find that tunneling stops when
\beq
\La_O \gsim |\la|^{4/3} \ .
\eeq
Thus, even for $|\la| \sim 10^{-60}$, although the effective
cosmological constant is reduced by a factor of $10^{20}$, one
membrane cannot suppress it to the observed value.

In general, however, it is important to note that several different
2-branes with various fundamental charge densities may be expected to
arise from different degenerating cycles.  Suppose that another brane
begins nucleating as the first membrane reaches its endpoint.  The
initial conditions for this new membrane are identical to those for
the first brane, except that now the role of the initial bare
cosmological constant is played by $\La_O \sim |\la|^{4/3}$. For
appropriate charge densities, $n$ branes may reduce the cosmological
constant to $|\la|^{(4/3)^n}$.  For $|\la| \sim 10^{-60}$, three
branes are sufficient to reduce the cosmological constant to its
observed value.

So far we have considered only the `conservative' $\be=3/2$ case.  For
larger values of $\be$ more complete relaxations of the cosmological
term are possible.  For general $\be$, a single membrane may relax the
cosmological constant to $\Lambda_O \gsim |\lambda|^{2 - \be^{-1}}$.
Thus, even for the $\be =2$ case, only two stages are required.  Note
also that in these multi-brane scenarios, in principle quite complex
dynamics can arise, with periods of slow relaxation interspersed with
more rapid changes.

\subsection*{Membrane tension and charge density}

So far we have taken the membrane tension, and especially charge density,
$\rho_2$, as freely adjustable (and very small!) parameters.  What is the
situation for these quantities in string theory?  In particular is there
a mechanism that allows, or even prefers, such small values?

In (9+1)-dimensions, the tension and charge density of type II D$p$-branes is
$T_p=\rho_p =\frac{2\pi}{\ell_s^{p+1}}$,
where the string length $\ell_s$ is defined in terms of the
fundamental string tension through
$T_{\rm F1}=1/(2\pi\alpha')=2\pi/\ell_s^2$. However we wish to know
the situation in 4-dimensions. There are two possibilities for 
how an effective 2-brane can arise in 4-dimensions.

First consider the case of a D2-brane in 10 dimensions 
descends directly to a 2-brane in 4 dimensions.  The 10-dimensional
supergravity action is compactified on a Calabi-Yau manifold with
string-frame volume $V_6$.  The physical effective tensions and charge
densities are then determined in 4-dimensional Einstein frame, with conventional 
normalizations of all kinetic terms.
For the 2-brane case of interest a short calculation leads
to the 4-dimensional effective charge density
\beq
\rho_2 \bigr|_{\rm 4D, eff} =
\frac{2\sqrt{2} \pi g_s^2}{M\ell_s^3} \ .
\label{rho2Eframe}
\eeq
where $g_s = e^{ \langle \phi \rangle}$ is the string
coupling.  The tension is $\tau_2 = M\rho_2/\sqrt{2}$.

Clearly, we can only obtain a sufficiently small charge density by
taking extreme values for $\ell_s$ and/or $g_s$.  For example, with
the canonical choice of string scale $\ell_s\simeq
(10^{17}\gev)^{-1}$, we find that a charge density of $\rho_2 \lsim
10^{-90}$ requires $g_s\lsim 10^{-44}$.  Alternatively we could take
$\ell_s$ to be a larger length scale.  These non-canonical cases
include the `large extra dimension' scenario with
$g_s\sim 1$ and some number of sub-millimeter dimensions.  However,
given the success of quantum field theory at colliders such as LEP and
the Tevatron, $\ell_s \lsim (\tev)^{-1}$ at the very best.  Although
an improvement, this still requires a tiny string coupling $g_s\lsim
10^{-23}$ to generate a sufficiently small $\rho_2$.  It is difficult
to understand how to accommodate the Standard Model in such an extreme
corner of string theory moduli space.

In principle a more promising alternative employs branes wrapped on
homology cycles of the compactification, and whose volume approaches
zero classically, as for conifolds~\cite{conifold}. Indeed wrapping on
such degenerating cycles leads to nearly tensionless branes, but
unfortunately small charge densities are not achieved.

Specifically, if a D$p$-brane of tension $T_p$ wraps a $k$-cycle $a_k$
of the compactification manifold, where $k \le p$ and the volume of
$a_k$ is ${\rm Vol}(a_k)$, then the result in the effective
4-dimensional theory is a $(p-k)$-brane of tension $\tau_{(p-k)} \sim
T_p \cdot {\rm Vol}(a_k)$.  If ${\rm Vol}(a_k)$ approaches zero, \ie,
$a_k$ is a degenerating cycle, a nearly tensionless object exists in
the 4-dimensional theory.  This is consistent with the
higher-dimensional quantization rules for brane properties.
Note, however, that as we wish to keep the Planck scale $M$ fixed,
these statements do not apply to the case where
a $(p-2)$-cycle of a factorizable compactification, such as a
$T^{p-2}$ in $K\simeq T^6$ or $K\simeq K3\times T^2$, degenerates.
Rather vanishing tension (with respect to $M$) applies to a
non-trivial CY 3-fold with degenerating cycles.  However a detailed
calculation shows that in this non-factorizable case the relation
between the tension and charge density picks up a dependence
on the size of the cycle, the end result being that the charge density
only logarithmically depends on the volume of the vanishing cycle (at least
for all cases studied).  Thus only very slightly smaller charge densities
than in the direct descent case are possible.  

Nevertheless, the phenomenon of classically vanishing tensions
arising from the wrapping of branes on degenerating cycles is
intriguing.  
One aspect that is worth noting is that the classical phenomenon of
true degeneration and corresponding
tensionless, or massless, states is typically not realized in the full
quantum theory.  Instead the effective (3+1)-dimensional 2-brane will
have a dynamically generated non-perturbative tension, which may be
exponentially small.  It is known that in some cases a tension
is generated from the dynamics of the would-be massless particle
states arising from a D2-brane wrapping the 2-cycle.  These particle
states realize a non-Abelian gauge theory, presumably in a sector
hidden with respect to the Standard Model, whose low-energy
non-perturbative dynamics can break supersymmetry. (This sector is
conceptually distinct from hidden sectors postulated to provide
supersymmetry breaking for the supersymmetric Standard Model.)  This
leads to a potential for the volume modulus of the cycle, which
stabilizes it at a scale $\La \sim \exp(-8\pi^2/b_1
g_{YM}^2)M$~\cite{mayr96,mayr00}.  Once the
cycle is stabilized at this small scale, membranes wrapping this cycle
have tension that are also proportional to $\La$
and thus can be very small, even for $g_{YM} \sim 1$. 

Thus, in summary, given our present understanding we just have to accept the
exceedingly small charge density $\rho_2$ as an input parameter
of a phenomenological model that awaits to be explained.

\subsection*{Summary and comments}

Any viable scenario for the solution of the cosmological
constant problem based on a relaxation principle must satisfy
two basic constraints that are in tension with one another:
the cosmological constant must relax sufficiently quickly from high
scales, but must be stable on cosmological time scales at its present
value.  The mechanism of Abbott and Brown and Teitelboim
unfortunately fails this test.
However, as I have outlined, the enhancement of
multi-step jumps due to large density of states factors, which
typically leads to large tunneling probabilities, provides a 
possible solution to this dilemma.  

I discussed two representative scenarios differing in the
treatment of the effective temperature entering the density of states
factor.  In the simplest scenario, with $T \sim T_O$, the
exponentially most probable transition, excluding extremely
short-lived universes, is to universes that are most nearly flat.  By
requiring that this new vacuum be sufficiently stable, a
non-trivial constraint for a mechanism of this kind was derived.
This constraint
provides an intriguing relationship, plausibly though not necessarily
satisfied in Nature, between the supersymmetry breaking scale and the
geometric mean of the present-day effective cosmological constant and
Planck scales.

Alternatively, an analogy with black holes suggests a richer dynamics,
in which flat space plays a distinguished role and tunneling to
anti-de~Sitter space is forbidden.  In contrast to the previous scenario,
the cosmological constant relaxes through a several jumps, roughly
following a geometric
series.  The constraint of stability limits the range over which the
cosmological constant may be relaxed by any given membrane.  However,
two or more types of branes with radically different scales may relax
the cosmological constant to within observational bounds, and appeal
to the anthropic principle may be avoided.

However, the saltation theory outlined above is clearly seriously incomplete.
Not only are there many questions of principle to be addressed, such as the
stopping mechanism, the correct form of the density of state enhancement,
and the resolution of the temperature ambiguity, and the origin
of the extremely small membrane coupling $\rho_2$, but there
are also phenomenologically important extensions that have not yet
been explored.  For example we have not attempted to incorporate saltation
into a realistic FRW cosmological model including matter.
Thus, in particular, we have not addressed the dynamics of relaxation
following a phase transition, with it's attendant potential problems
(and opportunities for inflation).
It is also of course important to explore the variety of observational
and experimental consequences of the possible existence of light membrane
degrees of freedom.

It is interesting to note that most
model building in string/M theory has been based, implicitly or
explicitly, on the paradigm of minimizing a potential, while
the saltatory mechanism suggests a
different principle, based on the dynamics of relaxation of the
cosmological term.   Moreover such a principle might prefer
complex, near-singular configurations, of the
type most favored by string model-builders. Finally it is 
intriguing that the density of states and effective temperature 
considerations suggest possible deep connections with the theory of
black holes.

\section*{Acknowledgements}
I wish to thank the organizers of CAPP2000 in Verbier for their
kind invitation to speak (as well as their patience).  I would
also like to thank J. Feng, N. Kaloper, S. Sethi, G. Starkman, M. Trodden,
and F. Wilczek for enjoyable collaboration on portions of the work reported here,
as well as N. Arkani-Hamed, A. Giveon, J. Maldacena, Y. Oz and H. Verlinde for
useful discussions.

%
\def\NPB#1#2#3{Nucl. Phys. {\bf B#1} #2 (#3)}
\def\PLB#1#2#3{Phys. Lett. {\bf B#1} #2 (#3)}
\def\PLBold#1#2#3{Phys. Lett. {\bf#1B} #2 (#3)}
\def\PRD#1#2#3{Phys. Rev. {\bf D#1} #2 (#3)}
\def\PRL#1#2#3{Phys. Rev. Lett. {\bf#1} #2 (#3)}
\def\PRT#1#2#3{Phys. Rep. {\bf#1} #2 (#3)}
\def\ARAA#1#2#3{Ann. Rev. Astron. Astrophys. {\bf#1} #2 (#3)}
\def\ARNP#1#2#3{Ann. Rev. Nucl. Part. Sci. {\bf#1} #2 (#3)}
\def\MPL#1#2#3{Mod. Phys. Lett. {\bf #1} #2 (#3)}
\def\ZPC#1#2#3{Zeit. f\"ur Physik {\bf C#1} #2 (#3)}
\def\APJ#1#2#3{Ap. J. {\bf #1} #2 (#3)}
\def\AP#1#2#3{{Ann. Phys. } {\bf #1} #2 (#3)}
\def\RMP#1#2#3{{Rev. Mod. Phys. } {\bf #1} #2 (#3)}
\def\CMP#1#2#3{{Comm. Math. Phys. } {\bf #1} #2 (#3)}
\relax
%
\newcommand{\journal}[4]{{ #1} {\bf #2}, #3 (#4)}
\newcommand{\hepth}[1]{{[hep-th/#1]}}
\newcommand{\hepph}[1]{{[hep-ph/#1]}}
\newcommand{\grqc}[1]{{[gr-qc/#1]}}
\newcommand{\astro}[1]{{[astro-ph/#1]}}
%

\end{document}